\documentclass[aps,prx,reprint,twocolumn,superscriptaddress,english]{revtex4}
\usepackage{amssymb}
\usepackage{graphicx}
\usepackage{amsmath}
\usepackage{amsfonts}
\usepackage[T1]{fontenc}
\usepackage[latin9]{inputenc}
\usepackage{array}
\usepackage{multirow}
\usepackage{color}
\usepackage{esint}
\usepackage{bm}
\usepackage{color}
\usepackage{bbm}
\usepackage{hyperref}
\usepackage{babel}
\usepackage{titlesec}
\usepackage{soul}

\titleformat*{\section}{\large\bfseries}
\titleformat*{\subsection}{\bfseries}
\titleformat*{\subsubsection}{\itshape}

\begin{document}


\global\long\def\ud{\mathrm{d}}
\global\long\def\ui{\mathbbm{i}}
\global\long\def\id{\mathbbm{1}}

\title{Observation of nodal-line semimetal with ultracold fermions in an optical lattice}

\author{Bo Song}
\thanks{These authors contributed equally to this work.}
\affiliation{Department of Physics, The Hong Kong University of Science and Technology,\\ Clear Water Bay, Kowloon, Hong Kong SAR}

\author{Chengdong He}
\thanks{These authors contributed equally to this work.}
\affiliation{Department of Physics, The Hong Kong University of Science and Technology,\\ Clear Water Bay, Kowloon, Hong Kong SAR}

\author{Sen Niu}
\thanks{These authors contributed equally to this work.}
\affiliation{International Center for Quantum Materials, School of Physics, Peking University, Beijing 100871, China}
\affiliation{Collaborative Innovation Center of Quantum Matter, Beijing 100871, China}

\author{Long Zhang}
\affiliation{International Center for Quantum Materials, School of Physics, Peking University, Beijing 100871, China}
\affiliation{Collaborative Innovation Center of Quantum Matter, Beijing 100871, China}

\author{Zejian Ren}
\affiliation{Department of Physics, The Hong Kong University of Science and Technology,\\ Clear Water Bay, Kowloon, Hong Kong SAR}

\author{Xiong-Jun Liu}
\email{xiongjunliu@pku.edu.cn}
\affiliation{International Center for Quantum Materials, School of Physics, Peking University, Beijing 100871, China}
\affiliation{Collaborative Innovation Center of Quantum Matter, Beijing 100871, China}
\affiliation{CAS Center for Excellence in Topological Quantum Computation, University of Chinese Academy of Sciences, Beijing 100190, China}
\affiliation{Institute for Quantum Science and Engineering and Department of Physics, Southern University of Science and Technology, Shenzhen 518055, China}

\author{Gyu-Boong Jo}
\email{gbjo@ust.hk}
\affiliation{Department of Physics, The Hong Kong University of Science and Technology,\\ Clear Water Bay, Kowloon, Hong Kong SAR}

\maketitle

{\bf Observation of topological phases beyond two-dimension (2D) has been an open challenge for ultracold atoms. Here, we realize for the first time a 3D spin-orbit coupled nodal-line semimetal in an optical lattice and observe the bulk line nodes with ultracold fermions. The realized topological semimetal exhibits an emergent magnetic group symmetry. This allows to detect the nodal lines by effectively reconstructing the 3D topological band from a series of measurements of integrated spin textures, which precisely render spin textures on the parameter-tuned magnetic-group-symmetric planes. The detection technique can be generally applied to explore 3D topological states of similar symmetries. Furthermore, we observe the band inversion lines from topological quench dynamics, which are bulk counterparts of Fermi arc states and connect the Dirac points, reconfirming the realized topological band. Our results demonstrate the first approach to effectively observe 3D band topology, and open the way to probe exotic topological physics for ultracold atoms in high dimensions.
}

The past decade has witnessed great progresses in
search for topological quantum phases, in particular the topological insulators~\cite{Qi:2011wt,Hasan:2010ku} and semimetals~\cite{Dirac1,Dirac3,Hasan2015,Ding2015,Ding2017} in solid state materials which commonly have
strong spin-orbit (SO) couplings.
Among the topological phases, a semimetal phase has gapless bulk nodes protected by symmetry and topology~\cite{Young:2012kz,monopole}. Particularly, the nodal-line semimetal has degenerate bulk quasiparticles extending 1D line~\cite{nodalline1,nodalline2}, and can serve as a parent phase to further realize exotic states including Weyl semimetals and topological insulators. Unlike the boundary modes of a topological matter which can be resolved with transport measurements or ARPES technique~\cite{Qi:2011wt,Hasan:2010ku}, the bulk topology is usually harder to detect. For nodal-line semimetals, the line-shape nodes of solids are embedded in the 3D band structure and their direct imaging could be impeded by the complexity of the system~\cite{Lou2018}.

Recently, considerable efforts have been made in ultracold atoms to explore synthetic SO couplings and topological quantum phases beyond natural conditions~\cite{Dalibard:2011gg,Goldman:2014bv,Zhai:2015hg,Zhang2018}.
A number of interesting phases have been reported in optical lattice experiments, including the Haldane model~\cite{Jotzu2014}, a minimal 2D SO coupled model~\cite{Wu:2016kv} for quantum anomalous Hall effect~\cite{Liu:2014hj}, a supersolid-like phase~\cite{Li:2016tp} and a 1D symmetry-protected topological state~\cite{Song:2017uf}. In particular, realizations of 2D SO couplings ignite enormous interests to explore high dimensional topological states with ultracold atoms~\cite{Xu:2016im,Xu:2016cz,Wang2016,He:2018up,Wang:2018ic}. However, to date only 1D and 2D topological phases were implemented in atomic systems, but no 3D topological states have been experimentally achieved. The great challenge is how to characterzie 3D band topology in atomic systems, which cannot be measured by standard momentum-space tomography or band mapping, as widely used for detecting 1D and 2D phases.

Here, we realize for the first time a 3D topological semimetal with nodal lines for ultracold fermions in an optical lattice, and successfully observe the nodal lines of the 3D topological band. The realization is based on 2D SO coupling proposed in an optical Raman lattice~\cite{Poon2018}, which forms a 2D Dirac semimetal in the $x$-$y$ plane, together with a 1D linear SO coupling along the free space of $z$ direction~\cite{Lin:2011hn,Wang:2011ul,Cheuk:2012id}. We develop a novel technique to detect the 3D band topology through a series of measurements of spin textures on the symmetric planes tuned by Zeeman splittings, which effectively reconstruct the 3D topological band. The detection technique is generically applicable to the systems with (emergent) magnetic group or similar symmetries, which cover various types of 3D topological phases.

\begin{figure}
\centering
	\includegraphics[width=9.0cm]{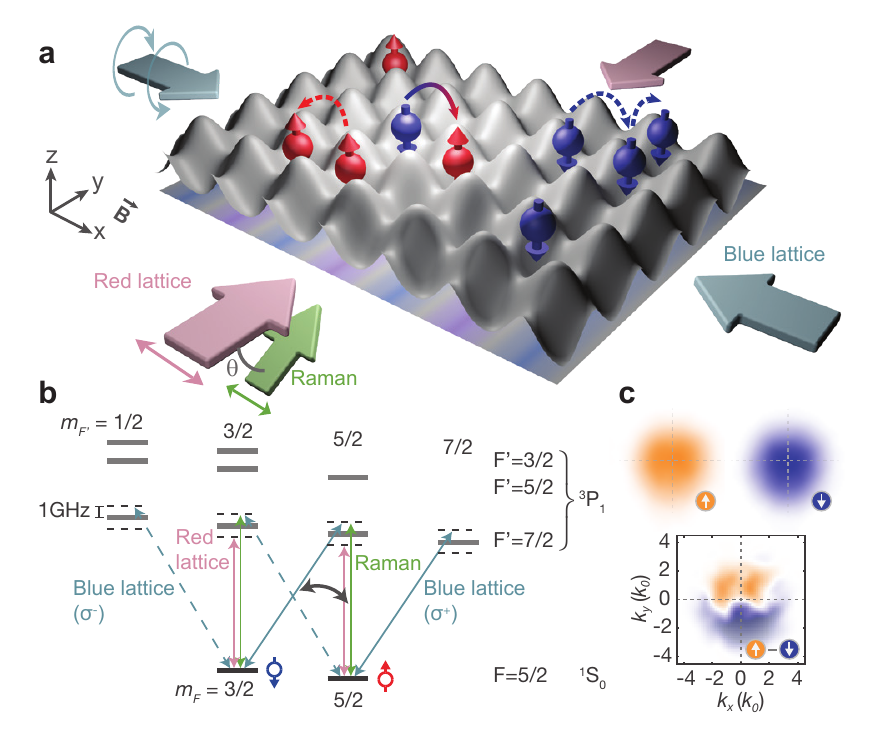}\\
	\caption{ (color online) {\bf Spin-orbit coupling in optical Raman lattices.} (a) A spin-dependent 2D optical lattice potential is created by two linearly-polarized beams, where spin-conserved and spin-flip hoppings occur. In addition to 2D SO coupling in the $x$-$y$ plane, linear SO coupling is induced in the $z$ direction by tilting the Raman beam. (b) The relevant electronic hyperfine levels of $^{173}$Yb atoms and intercombination transitions are shown. (c) The term $\cos{k_0 x}$ in Raman potential induces symmetric momentum distribution in the $x$ direction. However, a running-wave Raman potential along the $y$ direction causes the asymmetric momentum distribution. The graph in (c) shows the typical differential momentum distribution $n_{{\uparrow}}(k_x,k_y)-n_{{\downarrow}}(k_x,k_y)$ obtained from  spin-selective images after TOF expansion. }	\label{fig1}
\end{figure}

We start with the experiment for a two-component degenerate $^{173}$Yb Fermi gas
of atom number $ N_{\uparrow,\downarrow} = 5\times 10^3$ and prepared at $T/T_F \simeq$~0.5, where $\lvert{\uparrow, \downarrow}\rangle=\lvert{m_F=5/2, m_F=3/2}\rangle$ represent hyperfine states of the ground manifold~\cite{Song:2016ep}. The spin quantization axis is precisely set along the $x$ direction by 8~G bias field, which minimizes unwanted multi-photon transitions. An optical AC Stark shift is induced to separate out an effective spin-1/2 subspace from other hyperfine levels [see Fig.~\ref{fig1}(b)]~\cite{Song:2016ep}. We utilize optical Raman lattices where periodic Raman potentials are imposed on and exhibit nontrivial relative symmetries with respect to the lattice~\cite{Liu:2014hj,Wu:2016kv,Wang:2018ic,Song:2017uf,Zhang2018}. A square optical lattice forms in the $x$-$y$ plane by two pairs of standing-wave lights, one blue- and one red-detuned from the principle resonant transition $F=$5/2$\to$$F'=$7/2, with the lattice depths denoted by $V_{x,\sigma}$ and $V_{y,\sigma}$ ($\sigma=\{\uparrow,\downarrow\}$) for the two lights, respectively [see Fig.~\ref{fig1}(b)]. The lattice potential depths are $V_{x,\sigma} \propto \sum_{F'} E_x^2/\Delta_{x,F'}$ and $V_{y,\sigma} \propto \sum_{F'} E_y^2/\Delta_{y,F'}$. Here $E_x (E_y)$ denotes the laser field propagating in the $x (y)$ direction and $\Delta_{x(y),F'}$ are the blue (red) detunings. In our setting, the lattice depth is spin-dependent as $V_{x,\uparrow}/V_{x,\downarrow}=$1.31 and $V_{y,\uparrow}/V_{y,\downarrow}=$0.69, calibrated by the modulation spectroscopy.
The system is free of lattice in the $z$ direction. Raman coupling is created between $\lvert{\uparrow}\rangle$ and $\lvert{\downarrow}\rangle$ by applying an additional blue-detuned plane-wave beam (green arrow in Fig.~\ref{fig1}(a)), polarized along $x$-axis and running in $y$-$z$ plane with an angle $\theta=$68$^{\circ}$ to $y$-axis~\cite{Poon2018}. The lifetime of atomic sample in the experiment is over $10$~ms after the optical Raman lattice potentials are fully ramped up.

\begin{figure}
\centering
	\includegraphics[width=9cm]{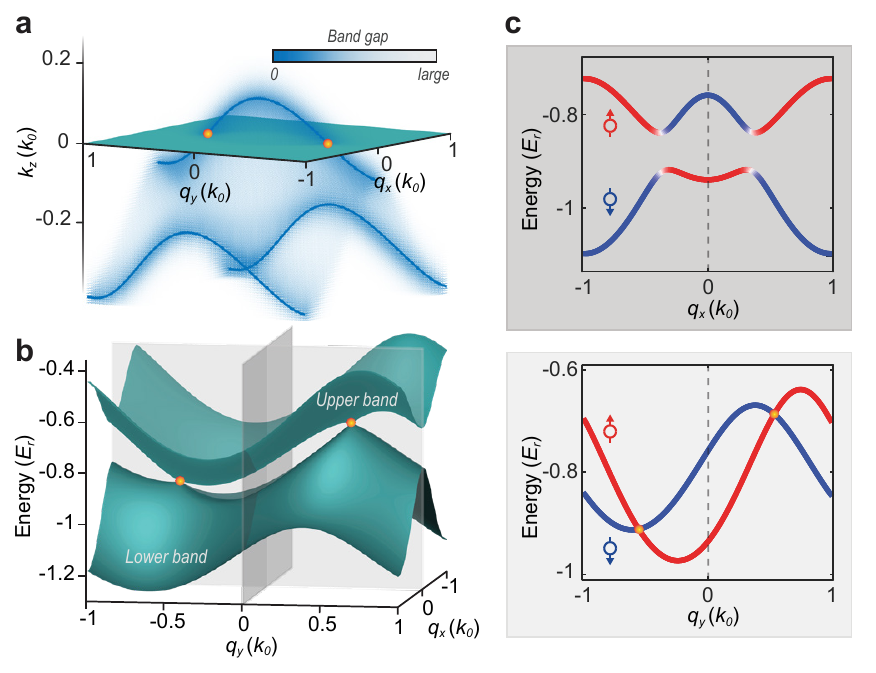}
	\caption{ (color online) \textbf {Nodal-line semimetal band structure}. (a) Numerical results for nodal lines formed in the 3D momentum space (solid lines), with the band gap being indicated by the color scale. At $k_z=0$ plane, for example, two Dirac points exist in the $q_x-q_y$ plane, giving a Dirac semimetal band structure in (b). In (c), the lowest two bands are shown along the $q_x$ direction ($q_y$=0) and $q_y$ direction ($q_x$=0). The parameters are taken based on experiment $\{V_{x,\uparrow},V_{x,\downarrow},V_{y,\uparrow},V_{y,\downarrow},M_R,m_z \}=\{6,4.17,4,5.22,0.5,0.13\} E_r$, where the recoil energy $ E_{r} =\hbar^2k^2_0/2m=h\times 3.735$~kHz with the Planck constant $h$, and $q_y$ is shifted as $q_y\rightarrow q_y+0.24\pi/a$ to ensure the positions of Dirac points symmetric. }
	\label{fig2}
\end{figure}

The Raman potential is the key ingredient to realize the topological semimetal, and is in the 3D form $V_R=M_R\cos{k_0 x}e^{ik_1y}e^{ik_2 z}\lvert{\downarrow}\rangle\langle{\uparrow}\lvert+{\mbox{H.C.}}$, where $k_1=k_0\cos\theta$, $k_2=k_0\sin\theta$, and $k_0=2\pi/\lambda$ for $\lambda=$556~nm. The total Hamiltonian realized in the experiment then reads
$H_{\rm total}=\vec p^2/2m+\sum_{\sigma=\uparrow,\downarrow}(V_{y\sigma}\cos{2 k_0 y}+V_{x\sigma}\cos{2k_0 x})+V_{R}+ m_z(\lvert{\uparrow}\rangle\langle{\uparrow}\lvert-\lvert{\downarrow}\rangle\langle{\downarrow}\lvert)$, where $m$ is the atom mass. The Zeeman term $m_z$ is controlled by the two-photon detuning $\delta$ as $m_z=(\delta-\delta_0)/2$, with $\delta_0$ the on-site energy difference between $\lvert{\uparrow}\rangle$ and $\lvert{\downarrow}\rangle$ atoms, depending on the lattice depth. We further confirm the value of the energy offset by monitoring the spin population in the optical Raman lattice, with which the Zeeman term $m_z$ is precisely determined. In Fig.~\ref{fig1}(c), the spin-dependent momentum shift of the atomic cloud is imaged experimentally in the $x$-$y$ plane after time-of-flight (TOF) expansion, showing a 2D pattern and revealing the 2D SO coupling in the lattice plane. Such 2D SO coupling leads to 2D Dirac semimetal for fixed $k_z$~\cite{Poon2018}, whereas the remaining plane-wave term $e^{ik_2z}$ in $V_R$ induces an additional 1D linear SO coupling, which modulates the 2D Dirac semimetals along $k_z$ axis to form into 3D nodal-line semimetal.

\begin{figure*}[tbp]
  \centering
	\includegraphics[width=16.1cm]{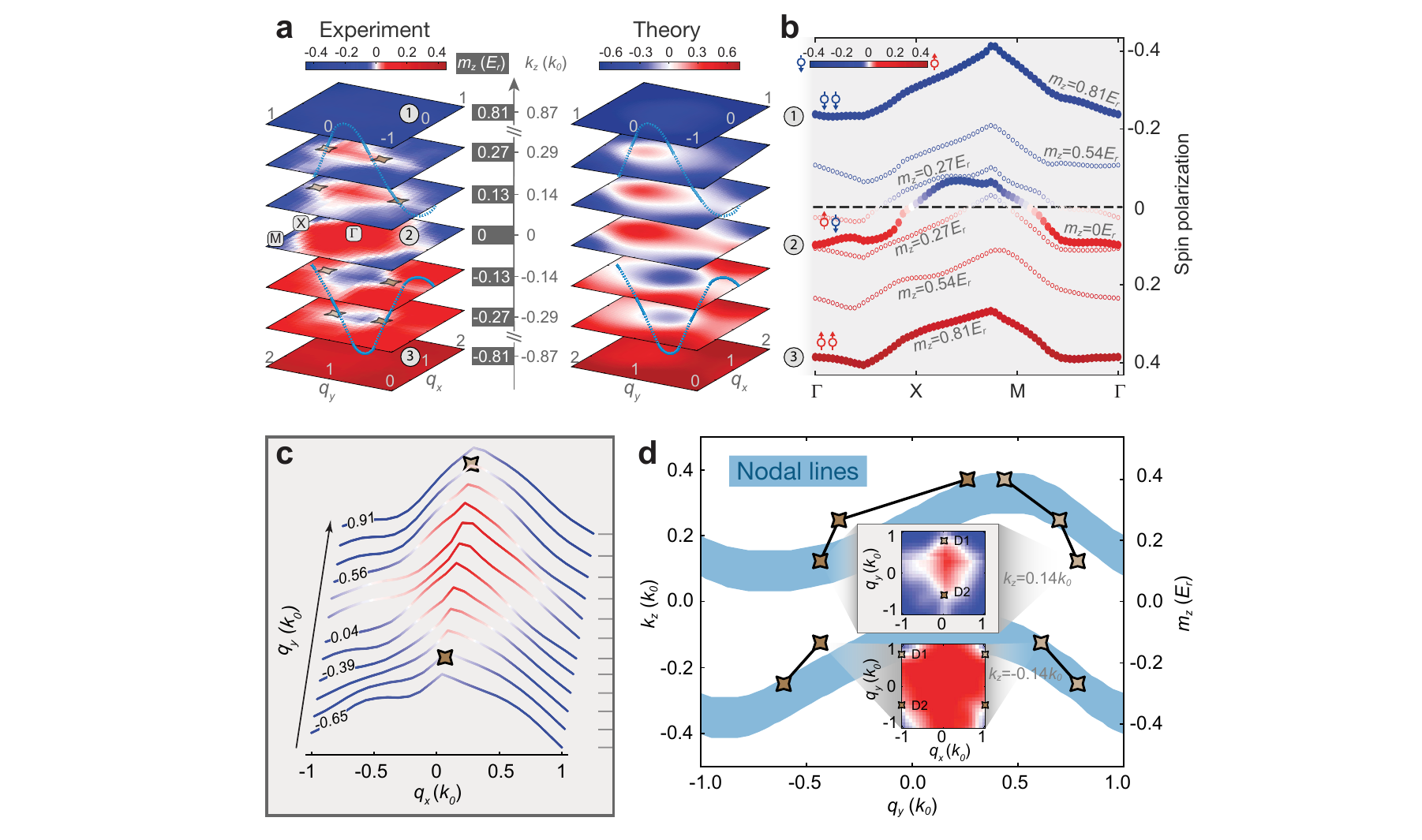}
	\caption{ (color online) \textbf{Measurement of nodal lines in the 3D momentum space}. (a) We map out 3D band topology by reconstructing $k_z$-resolved spin textures from the $k_z$-integrated spin texture obtained at different Zeeman splitting $m_z$. The measured Dirac points (yellow star) in the experiment, with the locations determined by the junctions of zero spin-polarization lines in $q_y$ direction, are consistent with prediction (blue curves) in the momentum space. The spin texture differs slightly from the prediction, and may ascribe to imperfect preparations which affect the position of zero spin-polarization. (b) Band inversion is experimentally observed around the $k_z=0$ plane (denoted by \textcircled{2}), when two bands are resonantly coupled by SO coupling. (c) From the spin textures measured experimentally in (a), the $q_y$-component of the nodal point in the $k_z$ plane can be determined by identifying the topological phase transition of the quasi-1D spin texture obtained at each $q_y$. (d) Finally, we can reconstruct nodal lines projected onto the $q_y$-$k_z$ plane. The measured nodal lines are consistent with the theoretical prediction within measurement uncertainty (light blue region). The parameters are $\{V_{x,\uparrow},V_{x,\downarrow},V_{y,\uparrow},V_{y,\downarrow},M_R \}=\{2.3(5),1.6(3),2.3(2),3.0(2),0.68(5)\}\times E_r$.}
	\label{fig3}
\end{figure*}

To understand the 3D band structure, we derive the Bloch Hamiltonian from $H_{\rm total}$ in the $s$-band regime (see supplementary material for details~\cite{SI})
\begin{eqnarray}\label{Bloch1}
H=[m_z+\frac{\hbar^2(k_z k_2)}{2m}+h_z^{2D}]\sigma_z+h_y^{2D}\sigma_y+h_0\sigma_0.
\end{eqnarray}
Here the SO coupling within the 2D lattice plane renders the quasi-momentum ($q_{x,y}$) dependent effective magnetic fields $(h_y^{2D},h_z^{2D})\propto(\sin q_x a,\cos q_x a+\cos q_y a)$ whose amplitudes are determined by Raman coupling strength, with $a=\pi/k_0$, and $\sigma_{x,y,z}$ the Pauli matrices on spin space, and the $h_0$-term is the spin-independent dispersion~\cite{SI}. The running-wave term $e^{ik_2z}$ of the Raman potential couples the spin-up (spin-down) states with the $z$-directional kinetic energy $\hbar^2(k_z\pm k_2/2)^2/2m$, giving the linear SO term $\frac{\hbar^2(k_z k_2)}{2m}\sigma_z$. Thus the full effect of the Raman coupling leads to the coupling between 3D momentum and two spin components ($\sigma_{y,z}$), which realizes the 3D nodal-line topological semimetal. 

For $k_z=0$, the above Hamiltonian describes a 2D Dirac semimetal with the Dirac points determined by $m_z=-h_z^{2D}$ and $h_y^{2D}=0$~\cite{Poon2018,SI}. In addition, the 1D linear SO term in the $z$ direction effectively modulates the Zeeman energy $m_z$, shifting the positions of the Dirac points at different $k_z$ layers. Therefore, the Dirac points at all $k_z$ layers form the nodal lines of the 3D topological semimetal. The numerical results using a plane-wave expansion are shown in Fig.~\ref{fig2}(a), where the nodal lines in a typical band structure are obtained~\cite{SI}. In Fig.~\ref{fig2}(b) the band structure in the $q_x$-$q_y$ plane at $k_z=0$ shows two Dirac points at $(q_x,q_y)=(0,\pm q_y^D)$ along the $\Gamma$-$X$ line, with an energy difference resulted from $h_0$ term and proportional to $\sin (q_y^Da)$. The 1D non-trivial spin texture is obtained in $q_x$ dimension for $|q_y|<q_y^D$, with the spin polarization changing sign at two momentum points between $\Gamma$ and $X$ [Fig.\ref{fig2}(b,c)]. Such two momenta are called band inversion points~\cite{ZhangL2018} which are related to a nonzero winding number~\cite{Song:2017uf,SI}. In contrast, the spin texture is trivial for $|q_y|>q_y^D$. The Dirac points at $q_y=\pm q_y^D$ then mark the transition between topological and trivial regimes for the quasi-1D bands along $q_x$ direction, and are characterized by the junctions of two band inversion lines formed by the band inversion points extending to $q_y$ direction in the $q_x$-$q_y$ plane. In the experiment, we set the lattice depths around 1$\sim$3$E_r$ to maximize the value of $q_y^D$ and the $z$-directional dispersion of nodal lines.

We proceed to experimentally map out the nodal lines by detecting Dirac points from spin textures with different $m_z$. The lattice potentials are adiabatically ramped to the final value at the constant two-photon detuning. Subsequently we record a spin-sensitive momentum distribution after TOF, and reconstruct a spin-texture of the lowest energy band with different $m_z$ in the $q_x$-$q_y$ plane by folding the atomic distribution into the first Brillouin zone~\cite{Song:2017uf}. During the expansion, we apply a resonant 556~nm light which selectively blasts a certain $m_F$ hyperfine state~\cite{Song:2017uf} followed by absorption imaging with 399~nm light resonant to the $^1$S$_0$-$^3$P$_1$ transition~\cite{SI}.

\begin{figure}
	\centering
	\includegraphics[width=8.7cm]{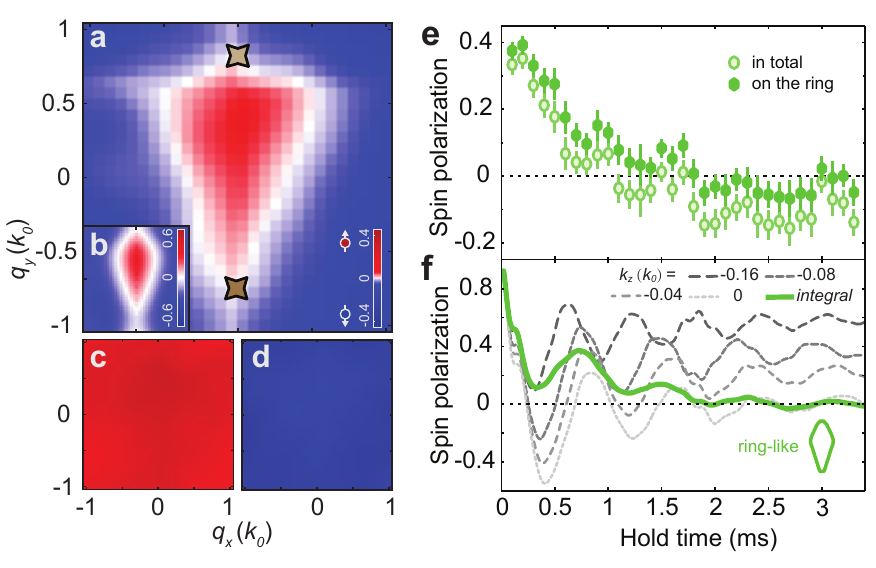}
	\caption{ (color online) {\textbf{Measuring band inversion lines from quantum quench dynamics}}. The spin-polarized gas initially prepared at $m_z=-4.6(2)E_r$ is quenched to a semimetal band [(a), $m_z=0.20(7) E_r$, and (b) numerical simulation with the same $m_z$] and trivial bands [(c), $m_z=-0.74(7)E_r$ and (d), $m_z=0.74(7)E_r$]. The time-averaged spin textures (a,c,d) are obtained by the evolution from $1~ms$ to $3~ms$ after quench. The band inversion lines, as characterized by zero time-averaged spin-polarizations, are observed only when the system is quenched to topological regime (a), consistent with the theoretical prediction in (b). (e) The evolution of spin polarizations integrated over momentum space and over the white ring are monitored at different hold times after the quench to the topological regime. (f) The observed decaying spin dynamics is consistent with theoretical prediction calculated based on the Lindblad equation with a decaying coefficient $\gamma=0.05$, chemical potential $\mu=-3.4E_r$, and temperature $T=0.42E_r$ (f)~\cite{SI}. Other numerical parameters are taken from the experiment. The integrated spin polarization over $k_z$ decays faster (solid green line) than that for each single $k_z$-plane (dashed grey lines), since the oscillations with different $k_z$ are not in phase. Error bars represent 1$\sigma$ standard error of the mean.}
	\label{fig4}
\end{figure}

Our key observation is that the nodal lines can be probed by effectively reconstructing the 3D band structure from the 2D momentum ($q_{x,y}$) resolved spin textures with different $m_z$. We reach this observation from two steps. First, the experimentally measured 2D spin texture for different $m_z$, with $k_z$ layers being integrated out, is in excellent agreement with the total spin texture obtained numerically with the same $m_z$ based on the 3D Hamiltonian $H_{\rm total}$ [compare the experimental with theoretical results in Fig.~\ref{fig3}(a)]. This measurement confirms the realization of the expected 3D semimetal in the current experiment. Second, the realized 3D semimetal phase has an intrinsic feature that the $k_z$-integrated spin texture for different $m_z$ is identical to the spin texture for different $k_z$ layer but fixed $m_z$ [both renders the same numerical plots in Fig.~\ref{fig3}(a)]. This feature is due to a novel mechanism explained below. As at the $k_z=0$ layer, the Bloch states with momenta on band inversion lines are given by the Hamiltonian $H_{q_{x0},q_{y0},0}=0\sigma_z+h_y^{2D}(q_{x0})\sigma_y$ and have zero $z$-component spin polarization. Further, for layers with $k_z\neq0$, the Bloch states at $q_{x,y}=q_{x0,y0}$ is governed by the Hamiltonian (see Supplementary Material~\cite{SI})
\begin{eqnarray}
H_{q_{x0},q_{y0},k_z}=\alpha k_2k_z\sigma_z+\alpha k_z^2\sigma_0+h_y^{2D}(q_{x0})\sigma_y,
\end{eqnarray}
where $\alpha=\hbar^2/2m$. The Hamiltonian satisfies a magnetic group symmetry ${\cal M}_zH_{q_{x0},q_{y0},k_z}{\cal M}_z^{-1}=H_{q_{x0},q_{y0},-k_z}$, where ${\cal M}_z=\sigma_xK$ is the product of time-reversal symmetry ($i\sigma_yK$) and mirror symmetry ($i\sigma_z$), with $K$ the complex conjugate. Thus the Bloch states at $\pm k_z$-momenta are degenerate but have opposite $z$-component spin polarizations, giving an important consequence that contributions to total spin polarization from $\pm k_z$ layers cancel out. Thus the total spin texture integrated over $k_z$ layers renders the spin texture on the magnetic-group-symmetric plane, i.e. the $k_z=0$ layer, as measured in the experiment. Finally, the spin texture on the symmetric plane is modulated by $m_z$. The spin texture for the $k_z=0$ layer by scanning $m_z$ is identical to that for fixed $m_z=m_0$ by scanning $k_z$, which follows the relation (after neglecting a constant): $H_{q_{x0},q_{y0},0,{\tilde m_z}}=H_{q_{x0},q_{y0},{\tilde k_z},0}$, with the equivalence of scanning parameters given by $\tilde m_z=\hbar^2 k_2\tilde k_z/2m$. With those two steps, we conclude that measuring the $k_z$-integrated spin texture by scanning $m_z$ confirms the experimental realization of the expected 3D topological semimetal, and simultaneously maps out effectively the 3D spin texture and nodal lines without necessity of additional measurements.
This result is valid in the presence of external trapping potential which keeps the magnetic group symmetry (see Supplementary Material~\cite{SI}), and not restricted by tight-binding condition. Fig.~\ref{fig3}(b) shows experimental results of spin polarization between symmetric momenta in the $q_x$-$q_y$ space. A nontrivial spin texture is observed around $m_z=0$, with two lowest energy bands being inverted and coupled by SO interaction. No band inversion occurs when two bands are far-detuned, so the observed spin textures are trivial.

To further confirm the nodal lines, we measure the projected nodal points onto the $q_y$-$k_z$ plane. For this we construct a quasi-1D spin texture along the $q_x$ direction, obtained at each $q_y$ from the 2D spin texture [Fig.~\ref{fig3}(c)]. The dimensional reduction allows us to consider $q_y$ as a parameter, and to measure the critical $q_y^D$ of topological transition for the reduced 1D band in the $q_x$ direction~\cite{Song:2017uf}. Then the $q_y^D$-position of the Dirac points can be measured versus $k_z$. Between the Dirac points, the quasi-1D spin texture exhibits non-trivial winding. Fig.~\ref{fig3}(d) shows the projected nodal lines onto the $q_y$-$k_z$ plane, in agreement with the theoretical prediction within the experimental uncertainty, and confirms the capability of resolving the 3D band topology.

We finally probe the far-from-equilibrium spin dynamics, following a quench from a deep trivial to topological regime, which allows to identify the band inversion lines and Dirac points with relatively higher resolution. On the band inversion lines, the complete spin-flip transitions occur due to the resonant coupling and lead to vanishing time-averaged spin polarization after quench~\cite{ZhangL2018}. In this measurement, we begin with a fully spin-polarized Fermi gas, adiabatically loaded into a trivial band, and suddenly change $m_z$ (i.e. two-photon detuning) to the semimetal regime. In Fig.~\ref{fig4}(a,c,d), we show experimentally the time-averaged spin textures in quench dynamics by monitoring spin evolution [Fig.~\ref{fig4}(e)]. The band inversion lines are clearly observed in Fig.~\ref{fig4}(a) from the zero time-averaged spin polarization measured over 1~ms$\sim$3~ms hold time [Fig.~\ref{fig4}(e)]. Ending at two Dirac points, the band inversion lines are the bulk counterparts of Fermi arc states of the semimetal. The measured time-averaged spin texture and quench dynamics are consistent with the numerical simulations given in Fig.~\ref{fig4}(b,f)~\cite{SI}. The spin dynamics are featureless when quenching the system to fully spin-polarized trivial bands as described in Fig.~\ref{fig4}(c,d).

Our results demonstrate the first realization and observation of a 3D SO coupled topological band with bulk nodal lines for ultracold fermions. We detect the bulk nodal lines by effectively reconstructing the 3D band topology through a series of measurements of $k_z$-integrated spin textures tuned by Zeeman splittings. This technique is symmetry-dependent and can be broadly applied to detecting any 3D topological states of the (emergent) magnetic group symmetry or similar symmetries, including the states realized in 3D optical lattices. This work brings important insights into simulating measurable nontrivial phases beyond 2D limit, and opens the way to explore high-dimensional topological quantum physics for ultracold atoms.

\paragraph*{\bf Acknowledgement}
We appreciate the valuable discussions with Lin Zhang. This work was supported by the Joint Research Scheme sponsored by the Research Grants Council (RGC) of the Hong Kong and National Natural Science Foundation of China (NSFC) (Project No. N-HKUST601/17 and  No. 11761161003). G.-B. J. acknowledges the support from the RGC and the Croucher Foundation through ECS26300014, GRF16300215, GRF16311516, GRF16305317, C6005-17G-A and the Croucher Innovation grants respectively. G.-B. J also thanks for a partial support (SSTSP grant) from HKUST. X.-J. L. acknowledges the support from the National Key R\&D Program of China (2016YFA0301604), NSFC (No. 11574008 and No. 11825401), and the Strategic Priority Research Program of Chinese Academy of Science (Grant No. XDB28000000).

\paragraph*{\bf Competing interests.} The authors declare that they have
no competing interests.

\paragraph*{\bf Data availability.} The data that support the findings of this study are available from the corresponding authors upon reasonable request.
\newpage
\clearpage

\renewcommand{\thesection}{M-\arabic{section}}
\setcounter{section}{0}  
\renewcommand{\theequation}{M\arabic{equation}}
\setcounter{equation}{0}  
\renewcommand{\thefigure}{M\arabic{figure}}
\setcounter{figure}{0}  

\section*{Methods}
\subsection*{Experimental method} 
\begin{figure*}
	\includegraphics[width=10cm]{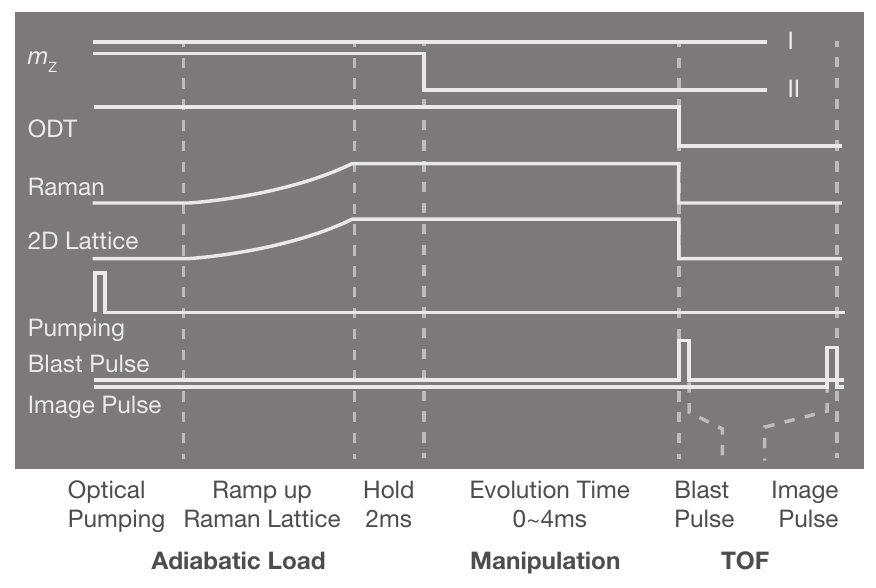}
	\caption{\textbf{Experimental sequence}. After evaporative cooling, atoms are adiabatically loaded into an optical Raman lattice by exponentially ramping up Raman lattice beams within 10~ms, followed by 2~ms hold. In the manipulation stage, the Zeeman energy $m_z$ is kept constant for spin-texture imaging in equilibrium (I), whereas suddenly changed for monitoring the quench dynamics (II). For the quench dynamics, a spin-polarized gas is prepared by optical pumping before the lattice ramp-up. Following a blast pulse, a spin-sensitive absorption image is taken after time-of-flight expansion.}
	\label{figM1}
\end{figure*}

\vspace{-0.1cm}
\paragraph*{ Experimental procedure} Experiments start with a degenerate Fermi gas of $^{173}$Yb atoms  in a far-detuned crossed optical dipole trap with the wavelength of 1064~nm, characterized by the trap frequencies of $\overline{\omega}=(\omega_x\omega_y\omega_z)^{1/3}=2\pi\times 126$~Hz. The 2D optical Raman lattice forms a 2D square optical lattice potential in the $x$-$y$ plane and consists of two pairs of standing-wave lights using the 556~nm intercombination transition. They are detuned from the principle resonant transition $F=$5/2$\to$$F'=$7/5 by +1.0~GHz (blue-detuned) and -1.31~GHz (red-detuned) respectively, with the lattice depth denoted by $V_{x,\sigma}$ and $V_{y,\sigma}$ ($\sigma=\{\uparrow,\downarrow\}$) for blue- and red-detuned beams respectively. In our setting, the lattice depth is spin-dependent as $V_{x,\uparrow}/V_{x,\downarrow}=$1.31 and $V_{y,\uparrow}/V_{y,\downarrow}=$0.69. The degenerate ground states of $^{173}$Yb atoms are lifted by both lattice beams, by which $m_F \leq \frac{1}{2}$ states are not affected by the Raman coupling. During the experiment,  other hyperfine levels $m_F \leq \frac{1}{2}$ can be safely neglected within the experimental resolution.  The quantization axis is precisely set along the $x$ direction by 8~G bias field, which minimizes unwanted multi-photon transitions. For the measurement of nodal lines, the $x$-$y$ lattice potential and the Raman beams are adiabatically ramped to the final value within 10~ms, during which the two-photon detuning is set to the final value. In a typical experimental setting, the lifetime of the atomic sample is at least 10~ms after the optical Raman lattice potentials are fully ramped up. The experimental sequence is shown in Fig.~\ref{figM1}.

\vspace{0.2cm}
\paragraph*{Optical Raman lattice potentials}
A 2D spin-dependent optical lattice is composed of two orthogonal blue-detuned (along the $x$ direction) and red-detuned (along the $y$ direction) standing-wave light field, $\textbf{E}_{x} = \textbf{E}_+ + \textbf{E}_-$ with $\textbf{E}_{\pm} = 1/\sqrt{2}\textbf{e}_{\pm} E_{x} \cdot \cos(k_0{x}+\omega t + \phi_{x}) $ and $\textbf{e}_{\pm} = 1/\sqrt{2} (\textbf{e}_y \pm i\textbf{e}_z)$, and $\textbf{E}_{y} = \textbf{e}_{x} E_{y} \cdot \cos(k_0{y}+\omega t + \phi_{y})$ respectively. The trap depth of the blue-detuned lattice is $V_{x,\sigma} = \sum_{F'} \hbar  \Omega^2_{x,\sigma,+,F'}/4\Delta_{x,F'} + \hbar \Omega^2_{x,\sigma,-,F'}/4\Delta_{x,F'} \propto \sum_{F'} E_x^2/\Delta_{x,F'}$ in which the Rabi frequency is $\Omega_{x,\sigma,\pm,F'} = \langle F, \sigma | e(y+z)/\sqrt{2} | F', m_F\pm1\rangle E_\pm$. Similarly, the red-detuned potential is $V_{y,\sigma} = \sum_{F'} \hbar \Omega^2_{y,\sigma,F'}/4\Delta_{y,F'} \propto \sum_{F'} E_y^2/\Delta_{y,F'}$ with $\Omega_{y,\sigma,F'} = \langle F, \sigma | ex | F', m_F\rangle E_y$. Here $\Delta_{x(y),F'}$ ($F'=3/2, 5/2$ and $7/2$) are the single-photon detuning of blue(red)-detuned light from the intercombination $| F = 5/2 \rangle \rightarrow | F'\rangle$ transitions. The Raman light field $\textbf{E}_{R} = \textbf{e}_x E_{R} \cdot e^{i(k_0y\cos\theta+k_0z\sin\theta+\omega t+\phi_{R})}$ is coupled with the blue-detuned lattice beam, generating a Raman potential $M_R = \sum_{F'} \hbar \Omega_{x,\sigma,+,F'}\Omega_{R,\sigma,F'}/4 \Delta_{x,F'} \propto \sum_{F'} E_xE_R/\Delta_{x,F'}$ with the Rabi frequency of the Raman light $\Omega_{R,\sigma,F'} = \langle F, \sigma | ex | F', m_F\rangle E_R$.

The lattice depth and the Raman coupling strength are calibrated by the modulation spectroscopy and the two-photon Rabi oscillation, respectively. The spin-dependent lattice potential introduces the on-site energy offset $\delta_0$ between $\lvert\uparrow\rangle$ and $\lvert\downarrow\rangle$ atoms, which is known from the lattice depth. We further confirm the value of the energy offset by monitoring the spin population in the optical Raman lattice, from which the Zeeman term $m_z=(\delta-\delta_0)/2$ is precisely determined.

\vspace{0.2cm}
\paragraph*{Spin texture imaging} To reconstruct a spin texture in the quasi-momentum space, we perform spin-resolved absorption imaging after a time-of-flight (TOF) expansion. At the beginning of the TOF expansion, we blast unwanted atoms in a spin component $\sigma$ by using a 556~nm light resonant to the $^{1}$S$_{0}(F=5/2) \rightarrow$$^{3}$P$_{1}(F'=7/2)$ transition. Three absorption images, $\mathcal{I} $, $\mathcal{I}_{\uparrow,\downarrow} $ and $\mathcal{I}_{\uparrow} $ are recorded using a 399~nm light resonant to the $ ^{1}$S$_{0} \rightarrow$$^{1}$P$_{1} (F'=7/2)$ transition where the subscript stands for the corresponding spin component removed by the blast light pulse. Finally, a momentum distribution of the atomic cloud for each spin, $\mathcal{D}_{\uparrow}^{TOF}(k_x,k_y)$ and $\mathcal{D}_{\downarrow}^{TOF}(k_x,k_y)$, are extracted from $ \mathcal{I}_{} - \mathcal{I}_{\uparrow} $ and $ \mathcal{I}_{\uparrow} - \mathcal{I}_{\uparrow,\downarrow} $ respectively. We note that a tiny fraction of atoms may occupy other spin states due to the imperfect isolation of the spin-$\frac{1}{2}$ subspace but our spin-sensitive detection is not susceptible to unwanted spin components.

The momentum distribution in the quasi-momentum $\mathcal{D}_{\sigma}(Q_x,Q_y)$ is constructed by folding $\mathcal{D}_{\sigma}^{TOF}(k_x,k_y)$ into the first Brillouin zone through shifting the integer number of $ 2k_0 $, $\mathcal{D}_{\sigma}(Q_x,Q_y) = \sum_{M,N}\mathcal{D}_{\sigma}^{TOF}(Q_x - 2Mk_0,Q_y - 2Nk_0)$ for $M,N \in Z$. Next, considering spin-momentum locking, we define states as $ \uparrow $, $\mathcal{D}_{\uparrow}(q_x,q_y) = \mathcal{D}_{\uparrow}(Q_x,Q_y)$ and $ \mathcal{D}_{\downarrow}(q_x,q_y) = \mathcal{D}_{\downarrow}(\text{mod}(Q_x-k_0,2k_0),\text{mod}(Q_y-\text{cos}\theta k_0,2k_0)) $, where $ \text{mod} $ is the modulo operator. Finally the spin texture $ P(q_x,q_y) $ is determined by $ (\mathcal{D}_{\uparrow}(q_x,q_y) - \mathcal{D}_{\downarrow}(q_x,q_y))/(\mathcal{D}_{\uparrow}(q_x,q_y) + \mathcal{D}_{\downarrow}(q_x,q_y))$.

In the spin texture, the values of spin polarization are averaged taking into account the finite optical resolution. The band inversion line with vanishing spin polarization is sensitive to non-ideal conditions that can affect the spin polarization in the experiment, including the thermal effects, imperfect loading procedure into the lowest energy band and the atom-atom interaction. We, however, reconfirm the band inversion lines from the far-from-equilibrium spin dynamics, which is less susceptible to those non-ideal conditions.

\begin{figure*}
	\includegraphics[width=14cm]{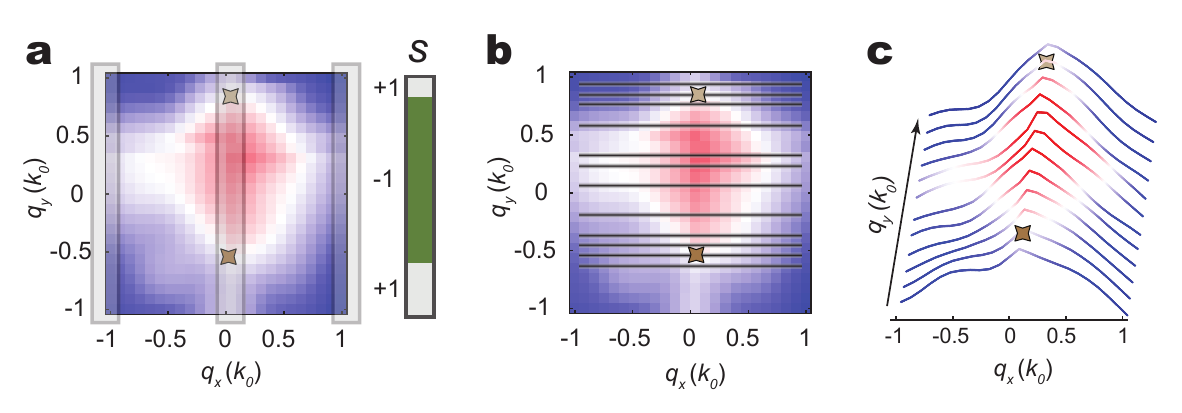}
	\caption{ \textbf{Determination of the Dirac points}. (a) In the first method, the quasi-momentum position of the Dirac point is determined by a sign product $ \mathcal{S} = sign(P(q_x=0,q_y) \cdot P(q_x=1k_0,q_y)) $. (b) In the second method, the Dirac point is measured as the spin-flip position. (c) Slice view of the spin texture of (b). The spin polarization varies along the $q_y$ direction.}
	\label{figS_Dirac_Point_Method}
\end{figure*}

\vspace{0.2cm}
\paragraph*{Determination of the position of the Dirac points}
We apply two methods to determine the momentum positions of the Dirac points from the result of spin texture measurement, illustrated in Fig.~\ref{figS_Dirac_Point_Method}.
First method is based on the topological phase transition points along the $ q_y $ direction in the spin textures. We first calculate the spin polarization along the $ q_x = 0$ and $ q_x = k_0 $ direction at different $ m_z $, $ P(q_x=0,q_y) $ and $ P(q_x = q_0,q_y) $ respectively. Next, we calculate a product of sign, $ \mathcal{S} = sign(P(q_x=0,q_y) \cdot P(q_x=1k_0,q_y)) $ (example shown in Fig.~\ref{figS_Dirac_Point_Method}(a)). To be noted, here the value of $ \mathcal{S} $ distinguishes different phases. Finally the positions of the Dirac points are determined by the sign-flip position $ q_{D1} $ along the $ q_y $ direction for each $ m_z $. Second method is based on the boundary between spin-$ \uparrow $ and $ \downarrow $ domain in the spin textures. Spin-flip positions $ q_{D2} $ along $ q_x = 0 $ and $ q_x = 1 k_0$, determine the locations of the Dirac points for $ m_z > 0 $ and $ m_z <0 $ respectively. The Dirac point position extracted from these two methods are consistent within the experimental uncertainty.

\subsection*{Theoretical method}\vspace{-0.1cm}

\paragraph*{Exact calculation of spin textures}
The total Hamiltonian realized in the experiment reads
\begin{eqnarray}
H_{\rm total}&=&\vec p^2/2m+\sum_{\sigma=\uparrow,\downarrow}(V_{y\sigma}\cos{2 k_0 y}+V_{x\sigma}\cos{2k_0 x})\nonumber\\
&&+V_{R}+ m_z(\lvert{\uparrow}\rangle\langle{\uparrow}\lvert-\lvert{\downarrow}\rangle\langle{\downarrow}\lvert),
\label{MethodHamiltonian1}
\end{eqnarray}
which can be diagonalized in the plane-wave bases. The Bloch states in the $n$-th band with lattice momentum $(q_x,q_y,k_z)$ take the form
\begin{eqnarray}
|\psi_{q_x,q_y,k_z,n}\rangle&=&[..., \phi_{q_x,q_y,k_z,n}(M,N,\uparrow), \nonumber\\
&&\ \phi_{q_x,q_y,k_z,n}(M,N,\downarrow), ...]^T.
\end{eqnarray}
Here $\phi_{q_x,q_y,k_z,n}(M,N,\uparrow)$ and $\phi_{q_x,q_y,k_z,n}(M,N,\downarrow)$ are superposition coefficients of the plane-wave bases $|q_x+2Mk_0,q_y+2Nk_0,k_z+k_2/2,\uparrow\rangle$ and $|q_x+k_0+2Mk_0,q_y-k_1+2Nk_0,k_z-k_2/2,\downarrow\rangle$, respectively, with $M,N$ being integers. The spin polarization $S_z(q_x,q_y,k_z,n)=\langle \psi_{q_x,q_y,k_z,n} |\hat{\sigma}_z |\psi_{q_x,q_y,k_z,n}\rangle$ of the Bloch state is given by
\begin{eqnarray}
		S_z(q_x,q_y,k_z,n)&=&\sum\limits_{M,N} \bigr[ |\phi_{q_x,q_y,k_z,n}(M,N,\uparrow)|^2\nonumber\\
&&- |\phi_{q_x,q_y,k_z,n}(M,N,\downarrow)|^2\bigr].
\end{eqnarray}

In thermal equilibrium the measured spin texture can be obtained from the density matrix $\rho$ with the matrix elements $\rho(q_x,q_y,k_z,n;q_x',q_y',k_z',n')$ written in the bases of Bloch states. The density of matrix in equilibrium takes the diagonal form $\rho(q_x,q_y,k_z,n;q_x,q_y,k_z,n)$. On the other hand, for the quench dynamics, we consider two cases. First, we take that the momentum distribution of $k_z$ has no decay and the density matrix elements take the form $\rho(q_x,q_y,k_z,n;q_x,q_y,k_z,n')$. Secondly, we consider decay for the momentum distribution of $k_z$. Then the density matrix elements take the form $\rho(q_x,q_y,k_z,n;q_x,q_y,k_z',n')$. The spin texture of a single $k_z$ layer is calculated by
\begin{equation}
	\begin{aligned}
		S_z(q_x,q_y,k_z)&=\frac{\text{Tr}_n(\hat{\sigma}_z\rho )}{\text{Tr}_n\rho },
	\end{aligned}
	\label{trace1}
\end{equation}
and the observable spin texture with $k_z$ being integrated out is calculated by
\begin{equation}
	\begin{aligned}
		S_z(q_x,q_y)&=\frac{\text{Tr}_{k_z}[\text{Tr}_{n}(\hat{\sigma}_z\rho )]}{\text{Tr}_{k_z}[\text{Tr}_{n}\rho ]}.
	\end{aligned}
	\label{trace}
\end{equation}
In numerics, the lowest five bands are taken into account in calculation for pre- and post-quench Hamiltonians.

\paragraph*{Equivalence between the $k_z=0$ layer and $k_z$-integrated spin textures}
We show analytically that the layer with $k_z=0$ and $k_z$-integrated spin textures are equivalent in that their band inversion lines are the same. Thus measuring the $k_z$-integrated spin texture yields the locations of Dirac points of the $k_z=0$ layer.
The Bloch Hamiltonian for $s$-band derived from Eq.~\eqref{MethodHamiltonian1} reads
\begin{eqnarray}
H=[m_z+\frac{\hbar^2(k_z k_2)}{2m}+h_z^{2D}]\sigma_z+h_y^{2D}\sigma_y+h_0\sigma_0.\label{methodHamiltonian2}
\end{eqnarray}
In the tight-binding regime the SO terms in the above Hamiltonian are obtained by
\begin{eqnarray}
h_0&=&2t_{y-}\cos{(q_ya+\phi)}-2t_{x-}\cos{q_xa}+\hbar^2k_z^2/2m,\nonumber\\
h_y^{2D}&=&2t_{so}\sin{q_xa},\\
h_z^{2D}&=&2t_{y+}\cos{(q_ya)}-2t_{x+}\cos{q_xa},\nonumber
\end{eqnarray}
where $t_{x\pm}$ and $t_{y\pm}$ are corresponding hopping coefficients along $x$ and $y$ directions, respectively, and $\phi$ is a constant~\cite{SI}. The Bloch Hamiltonian at the spin-balanced momenta $(q_{x0},q_{y0})$ with $m_z=-h_z^{2D}(q_{x0},q_{y0})$ and for the $k_z=0$ layer reads
\begin{equation}
	\begin{aligned}
		H_{q_{x0},q_{y0},0}=0\sigma_z+h^{2D}_y(q_{x0})\sigma_y,
	\end{aligned}
\end{equation}
where the irrelevant terms have been discarded. Further, the Bloch Hamiltonian at $(q_{x0},q_{y0},k_{z})$ points should be
\begin{equation}
	\begin{aligned}
		H_{q_{x0},q_{y0},k_{z}}= \alpha\sin \theta k_0k_z\sigma_z+\alpha k_z^2\sigma_0+h^{2D}_y(q_x)\sigma_y.
	\end{aligned}
\end{equation}
The above Bloch Hamiltonian satisfies an emergent magnetic group symmetry defined by
\begin{equation}
	\begin{aligned}
		{\cal M}_zH_{q_{x0},q_{y0},k_z}{\cal M}_z^{-1}=H_{q_{x0},q_{y0},-k_z},
	\end{aligned}
\end{equation}
where ${\cal M}_z=\sigma_xK$. As a result, the Bloch states at $\pm k_z$-momenta are degenerate but have opposite $z$-component spin polarizations. Thus after $k_z$ is integrated the polarization will be kept zero at $(q_{x0},q_{y0})$ momentum. The zero spin polarization momenta $(q_x,q_y)$ for the $k_z=0$ layer are still zero polarized after $k_z$ is integrated.


\paragraph*{Equivalence between scanning $k_z$ and scanning $m_z$}
This equivalence is generic, not relying on the tight binding regime. Note that in the Bloch Hamiltonian $H$ given in Eq.~\eqref{methodHamiltonian2} the term $h_z^{2D}$ is independent of $k_z$,
one can find immediately that
\begin{equation}\label{equivalence2}
		H _{q_x,q_y,k_z,0}= H _{q_x,q_y,0,\tilde m_z}+\frac{\tilde m_zk_z}{k_2}\sigma_0,
\end{equation}
where $\tilde m_z=\hbar^2k_2 k_z/2m$. The last constant $\sigma_0$ term can be discarded as an effective chemical potential. Finally we reach
\begin{equation}
		H _{q_x,q_y,k_z,0}= H _{q_x,q_y,0,\tilde m_z}.
\end{equation}
The above equivalences are clearly not restricted to tight binding regime, since the emergent magnetic group symmetry and the relation in Eq.~\eqref{equivalence2} are independent of the dispersion in the $x-y$ plane, as confirmed in numerical results. More numerical details can be found in Supplementary Material~\cite{SI}.

\begin{figure*}[htb]
	\centering
	\includegraphics[width=4in]{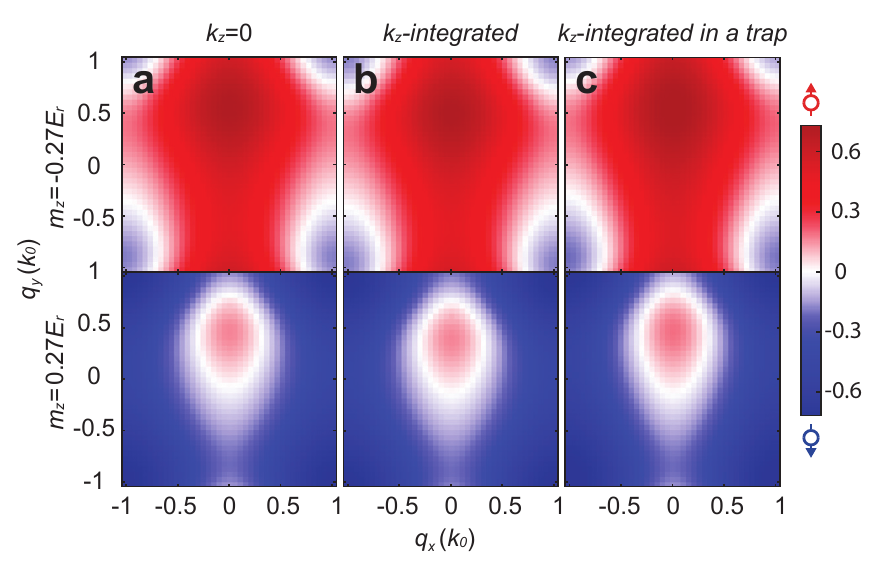}
	\caption{{\bf Theoretical static spin textures}
	(a) Spin textures are numerically calculated using plane-wave expansion for two different values of $m_z$ within the $k_z$=0 plane. The band inversion (zero spin polarization) lines are not affected in (b) $k_z$-integrated spin textures and (c) a harmonic trap based on experiment. The parameters are $(V_{x\uparrow},V_{x\downarrow},V_{y\uparrow},V_{y\downarrow},M_R,m\omega_z^2a^2,T)=(2.3,1.6,2.3,3.0,0.68,0.004,0.42)E_r$, with $\omega_z$ the trapping frequency along $z$ axis.}
	\label{figM3}
\end{figure*}

\begin{figure*}
	\centering
	\includegraphics[width=4in]{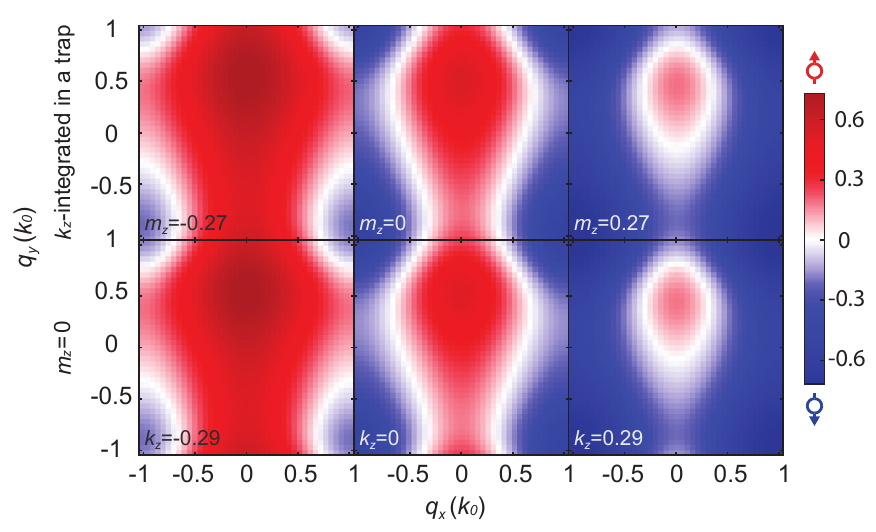}
	\caption{{\bf Equivalence between scanning of $k_z$ and scanning of $m_z$ with trapping} The $k_z$ integrated spin textures for variable $m_z$ with external trap (upper row) are equivalent to spin textures at fixed $m_z$=0 but in different $k_Z$ layers (lower row). The parameters are the same as in Fig.~\ref{figM3}.}
	\label{figM4}
\end{figure*}

\paragraph*{Effect of trapping}
In the current experiment a spin-independent external trapping $V_{\rm ex}(\bold r)$ is applied, in which case the Hamiltonian becomes $\tilde H=H_{q_{x0},q_{y0},k_z}+V_{\rm ex}(\bold r)$. It can be seen that under the transformation by ${\cal M}_z$ the trapping potential $V_{\rm ex}(\bold r)$, which is real, is unchanged. As a result, the total Hamiltonian $\tilde H$ still preserves the emergent magnetic group symmetry. Note that the (quasi)momentum is no longer good quantum number due to trapping potential, but the spin texture can be defined by projecting the eigenstates onto the (quasi)momentum space which is actually resolved in the TOF imaging. The spin polarizations contributed from the projected $\pm k_z$ layers again cancel out due to the symmetry and the conclusion is not affected. This implies that the mixing between different $k_z$ due to trapping cannot affect the integral result of the spin texture over all the $k_z$-momentum, as confirmed numerically in Fig.~\ref{figM3}.
Further, the shift of the projected $k_z$ by $m_z$ in the Bloch Hamiltonian is clearly not affected by the trapping, and the equivalence in Eq.~\eqref{equivalence2} is again valid. Thus scanning $m_z$ can still exactly map out the spin texture for each projected $k_z$ layer (equivalent to the $k_z$ layer without trapping), as shown by numerical calculation [Fig.~\ref{figM4}].
In this way, with our approach we can still reconstruct the 3D spin texture.
More details of the proof and numerical confirmation of the result are further given in Supplementary Material~\cite{SI}.

\newpage

%


%
%


\end{document}